\documentclass[preprint,review,12pt]
{elsarticle}

\usepackage{amssymb}
\usepackage{amsmath}
\usepackage{booktabs}
\usepackage{subcaption}
\usepackage{url}
\usepackage{xcolor}
\usepackage{enumitem}
\usepackage{amsthm}
\usepackage{algorithm}
\usepackage{algpseudocode}
\usepackage{multirow}
\usepackage{comment}

\newtheorem{definition}{Definition}

\journal{Information Systems}

\begin{document}

\begin{frontmatter}
\title{Compressed Boolean Matrix Multiplication\\ using Bicliques}

\author[1]{Nicolás Araya}
\ead{naraya2018@udec.cl}

\author[1]{José Zagal}
\ead{jzagal2018@inf.udec.cl}

\author[1]{Cecilia Hernández\corref{cor1}}
\ead{cecihernandez@udec.cl}

\author[2]{Gonzalo Navarro}
\ead{gnavarro@dcc.uchile.cl}

\cortext[cor1]{Corresponding author}

%% Author affiliation
\affiliation[1]{organization={Department of Computer Science, Universidad de Concepción},%Department and Organization
            addressline={Edmundo Larenas 219},
            city={Concepción},
            postcode={4070409},
            state={Bío-Bío},
            country={Chile}}
\affiliation[2]{organization={Millennium Institute for Foundational Research on Data (IMFD) and Department of Computer Science, University of Chile},%Department and Organization
            addressline={Beauchef 851},
            city={Santiago},
            %postcode={9050000},
            %%state={Santiago},
            country={Chile}}

\begin{abstract}

Boolean matrix multiplication is a fundamental operation in many graph-based applications, including web and social network analysis, information retrieval, and graph query evaluation. However, conventional sparse representations such as \textit{CSR}/\textit{CSC} still incur significant space and time costs on large graphs. In this work, we introduce a method for multiplying Boolean matrices based on a biclique-based compressed representation that decomposes them into dense bipartite subgraphs, thereby capturing large sets of implicit edges. We first improve the iterative biclique extraction algorithm of Hernández and Navarro by introducing percentile-based parameter adaptation that dynamically adjusts its parameters according to the degree distribution, enhancing both extraction efficiency and compression ratio. We then extend the Schoor multiplication algorithm to operate directly on biclique-encoded matrices, thereby enabling a compositional form of multiplication where the output can also be represented as a collection of bicliques. Our experimental results on large real-world web and social graphs show that our approach achieves substantial reductions in space and computation time with respect to \textit{CSR}/\textit{CSC}-based representations. For example, using biclique extraction just as a tool to speed up multiplication (but retaining the \textit{CSR}/\textit{CSC} format) yields up to 34-fold speedups. Further, using the biclique-based representation for input and output yields up to a 10-fold reduction in space {\em and} up to a 290-fold speedup for matrix multiplication. These results demonstrate that our representation based on bicliques is both space- and time-effective to support an algebra of Boolean matrices.
\end{abstract}


\end{frontmatter}

% \linenumbers

\section{Introduction}

Graphs play a fundamental role in modeling a wide range of problems in many areas, including social networks \cite{Wu25}, knowledge graphs \cite{KG21}, the Web \cite{Hog20}, and scientific applications such as gene sharing \cite{watson2019methodology} or de Bruijn graphs \cite{Etz24}. A common representation for graphs is adjacency matrices, which enable algebraic operations between graphs using Boolean matrix arithmetic: transposing matrices, Boolean sums, matrix-vector and matrix-matrix multiplication, and transitive closures are fundamental in a wide range of tasks, including context-free grammar parsing and query processing in graph databases, multi-hop reachability computation, diffusion or influence analysis in social networks, and dynamism in evolving graphs.

Most of those applications involve {\em sparse} graphs, which means that their adjacency matrices have comparatively few non-zeros. 
The best known representations for sparse matrices include the classic Compressed Sparse Row (\textit{CSR}) format, which stores non-zero elements in row-major order, along with column indices and the starting position of each row. The Compressed Sparse Column (\textit{CSC}) format is the analogous structure for indexing columns. The space required by these representations is proportional to the number of edges in the graph. Further, they operate on sparse matrices much faster than the standard dense matrix representation. 

On large sparse graphs and matrices, however, the space and operation time remains and issue, especially on the most costly operations like matrix multiplication and transitive closure. In this paper we focus on the former operation, as it appears more frequently in applications.
One of the best practical sparse matrix multiplication algorithms is due to Schoor \cite{Schoor82}, which has an average time complexity of $O(C_xC_y/N)$ for multiplying matrices $X \times Y$ of dimensions $M \times N$ and $N \times L$, where $C_x$ and $C_y$ are the number of nonzero elements in $X$ and $Y$, respectively, assuming nonzero entries distribute uniformly in the matrices.
Recently, Arroyuelo et al.~\cite{arroyuelo2025evaluating} developed a compact-space implementation of Schoor's algorithm with an average time complexity of $O(C_xC_y\log(N)/N)$ on top of \textit{CSR}/\textit{CSC} representations.

Several schemes obtain further compression upon conventional sparse-matrix formats like \textit{CSR} and \textit{CSC}, by exploiting sparsenes and clustering~\cite{brisaboa2009k2}, similarity of adjacency lists~\cite{boldi2004webgraph,ferragina2022improving}, or clique and biclique patterns~\cite{hernandez2014compressed,glaria2021compact}, among others. Some of those formats enable efficient matrix-vector~\cite{francisco2022graph,ferragina2022improving} and matrix-matrix~\cite{arroyuelo2025evaluating} multiplication. In this paper we are interested in the compression based on bicliques~\cite{hernandez2014compressed}, which capture highly connected regions that frequently occur in social and web graphs, where communities or connectivity clusters induce high edge density. By compactly representing those dense substructures, compression methods represent many edges implicitly. Further, it has been shown that those compact representations can be used to speed up matrix-vector multiplication, where the matrix is sparse but the vector is dense~\cite{francisco2022graph}.

In this work we improve the biclique-based representation and extend its functionality to support Boolean matrix-matrix multiplications directly on the compressed matrices, improving both space usage and computation time compared to the state of the art. More precisely, our main contributions follow.

\begin{enumerate}
    \item We improve our previous biclique extraction~\cite{hernandez2014compressed} by defining an adaptive method that dynamically adjusts the parameters according to the degree distribution of the input graph. This adaptive strategy improves both compression ratio and the runtime, particularly for large real-world networks with heterogeneous degree distributions.
    \item We show that Schoor's algorithm~\cite{Schoor82} adapts naturally to biclique-based representations, and develop a version that directly operates over our representation. In particular, our version of Schoor's algorithm both receives the input and produces the output in the form of bicliques.
    \item We provide an extensive experimental evaluation on large real-world web graphs, comparing space usage and running time against \textit{CSR}/\textit{CSC}-based Schoor and other compressed representations that support matrix-matrix multiplications~\cite{arroyuelo2025evaluating}. In a first scenario, we retain the \textit{CSR}/\textit{CSC} format and use biclique extraction and multiplication just as a speedup, becoming up to 34 times faster than Arroyuelo et al.'s \textit{CSR}/\textit{CSC} baseline multiplication. In a second scenario, we use bicliques as the standard matrix representation and exploit the compositionality of our biclique-based Shoor's multiplication to maintain the matrices compressed with bicliques all of the time. We obtain space reductions up to 10-fold and time improvement factors of up to 290 with respect to the baseline. Our compression based on bicliques is shown to perform even better on the output of multiplications, which makes the method even stronger.
\end{enumerate}

\section{Related Work}

Efficient Boolean matrix multiplication has been extensively studied, though not so much for sparse matrices. One of the most efficient ones in practice is Schoor's~\cite{Schoor82}, which works on a {\em CSR}/{\em CSC}-like representation. We describe it in detail in Section~\ref{sec:schoor}.

In the context of graph structures, various space-efficient compressed representations have been proposed, which also support efficient computation of queries. Some of the most relevant methods include the Webgraph framework~\cite{boldi2004webgraph} and the $k^2$-tree~\cite{brisaboa2009k2}. Over the years, the Webgraph framework has been improved by exploiting locality, similarity, and node ordering in large Web and social graphs. On the other hand, $k^2$-tree exploits the sparseness and clustering of zeros of adjacency matrices. Other methods use other regularities; for instance,  Hernández and Navarro~\cite{hernandez2014compressed} exploit dense subgraphs that include cliques and bicliques to represent edges implicitly via their vertex sets, and on a later work Glaria et al.~\cite{glaria2021compact} showed how to compress undirected graphs by clustering maximal cliques. Pinar and Heath~\cite{pinar1999improving} used a compressed form to represent the matrix using the \textit{CSR} format.

Compressed graph representations also provide support for computing fast matrix operations. For instance, Francisco et al. \cite{francisco2022graph} reduced the matrix-vector multiplication time (where the matrix is sparse and the vector is dense) using the Webgraph framework of Boldi and Vigna~\cite{boldi2004webgraph} and the biclique extraction approach of Hernández and Navarro~\cite{hernandez2014compressed}. They evaluated the impact in the PageRank computation using an approximation of the power method~\cite{francisco2022graph}. Ferragina et al.\  \cite{ferragina2022improving} present a different approach to matrix-vector multiplication computation of non-binary matrices. They focus on grammar-based compression, emphasizing the similarities in the matrix more than its sparseness. They represent the matrix using an extension of \textit{CSR} that also stores (nonzero) cell values, and implement a grammar-based compression similar to RePair~\cite{larsson2000off}. The resulting representation allows matrix-vector operations in time proportional to the number of replacement rules generated by the compression. Nelson et al.~\cite{nelson2017queryable} propose matrix-vector and matrix-matrix multiplications using compressed Boolean matrices with compressed binary tree algorithms~\cite{nelson2019billion} and an adapted encoding algorithm and node reordering to better capture similarities of contiguous adjacency lists.

\section{Schoor's Multiplication Algorithm}
\label{sec:schoor}

\subsection{Original Algorithm}

Schoor~\cite{Schoor82} proposed one of the most efficient sparse matrix multiplication algorithms, over a {\em CSR}/{\em CSC}-like representation. The algorithm uses $5C + M + N$ words of space for an $M\times N$ matrix with $C$ nonzero cells. Its average time to compute $ Z = X \times Y$ is $O(C_xC_y/N)$, where $C_x$ and $C_y$ are the number of non-zero cells in matrices $X$ (of size $M \times N$) and $Y$ (of size $N \times L$), assuming the nonzero entries are uniformly distributed. The core idea is to intersect the set of nonempty columns of $X$ with the set of nonempty rows of $Y$. For each $1 \le t \le N$ in that intersection, every nonzero cell $X_{it}$ of $X$ and every nonzero cell $Y_{tj}$ of $Y$ contribute $X_{it} \cdot Y_{tj}$ to $Z_{ij}$. On Boolean matrices, this means that $t$ contributes a biclique to $Z$ formed by all the nonzero rows $i$ of column $t$ of $X$ on one side, and all the nonzero columns $j$ of row $t$ of $Y$ on the other. 

Recently, Arroyuelo et al.~\cite{arroyuelo2025evaluating} implemented Schoor's algorithm in a more compact form. Since the algorithm needs $X$ in {\em CSC} form and $Y$ in {\em CSR} form, they store both, yet using only $2(C+N+M)$ words and also compacted the intermediate space for computation. Their average computation time, in exchange, raises to $O(C_xC_y\log(N)/N)$.

\subsection{A Compact Implementation}

Arroyuelo et al.~\cite{arroyuelo2025evaluating} present an adaptation of Schoor's algorithm using the $\textit{CSR}$ and $\textit{CSC}$ matrix representations. This implementation improves space efficiency and achieves an average time complexity of $O(C_xC_y\log (N)/N)$.
The implementation uses integer arrays for the $\textit{CSR}$ format to store the non-empty rows (\texttt{rowId}), the non-empty columns for each row (\texttt{colsIdx}), and the starting position of each row (\texttt{rowPtr}). Similarly, for $\textit{CSC}$ it uses \texttt{colId}, \texttt{rowsIdx} and \texttt{colPtr} arrays. Arrays \texttt{rowPtr} and \texttt{colPtr} have a terminator entry that points to the first cell past the arrays \texttt{colsIdx} and \texttt{rowsIdx}.

The matrix multiplication computes $Z = X \times Y$ using $X$ in $\textit{CSC}$, $Y$ in $\textit{CSR}$, and $Z$ in $\textit{CSR}$ (plus $\textit{CSC}$, generated from $\textit{CSR}$), in two main steps:

\begin{description}
 \item[STEP 1: Intersections and Cartesian product sizes]

This step identifies intersections between columns of $X$ (in \texttt{colId}) and rows of $Y$ (in \texttt{rowId}) that participate in the multiplication. The intersection algorithm uses exponential or linear search depending on array sizes, finding all $t$ such that a column $t$ in $X$ matches a row $t$ in $Y$.
Each intersection $t$ enables the computation of the Cartesian product between cells $X_{it}$ of rows $i$ and cells $Y_{tj}$ of columns $j$. The cardinality of these products determines the total number of edges, though duplicates may occur. Rows $i$ are retrieved from \texttt{colPtr} and \texttt{rowsIdx} of $X$, and columns $j$ from \texttt{rowPtr} and \texttt{colsIdx} of $Y$. 
A structure called \texttt{Task} stores pointers and sizes of Cartesian products, which are collected in an array.

\item[STEP 2: Cartesian product computation]

This step computes the Cartesian products to identify the nonzero cells of the matrix $Z$. Since the same cell may appear in multiple products, two min-heaps are used to avoid duplicates: one for rows and one for columns.

All Cartesian products are first inserted in \texttt{HeapByRow}, which is sorted by the smallest row $i$ in the product. Elements are then extracted until all those sharing the smallest row are obtained. As rows are extracted, their columns $j$ are inserted in increasing order into \texttt{HeapByCol}. If the products have more than one row, they are reinserted into \texttt{HeapByRow} with the first row removed. The elements extracted from \texttt{HeapByCol} are used to directly produce $Z$ in $\textit{CSR}$ format.

\end{description}

\begin{figure}[tb]
  \centering
  \includegraphics[width=1\linewidth]{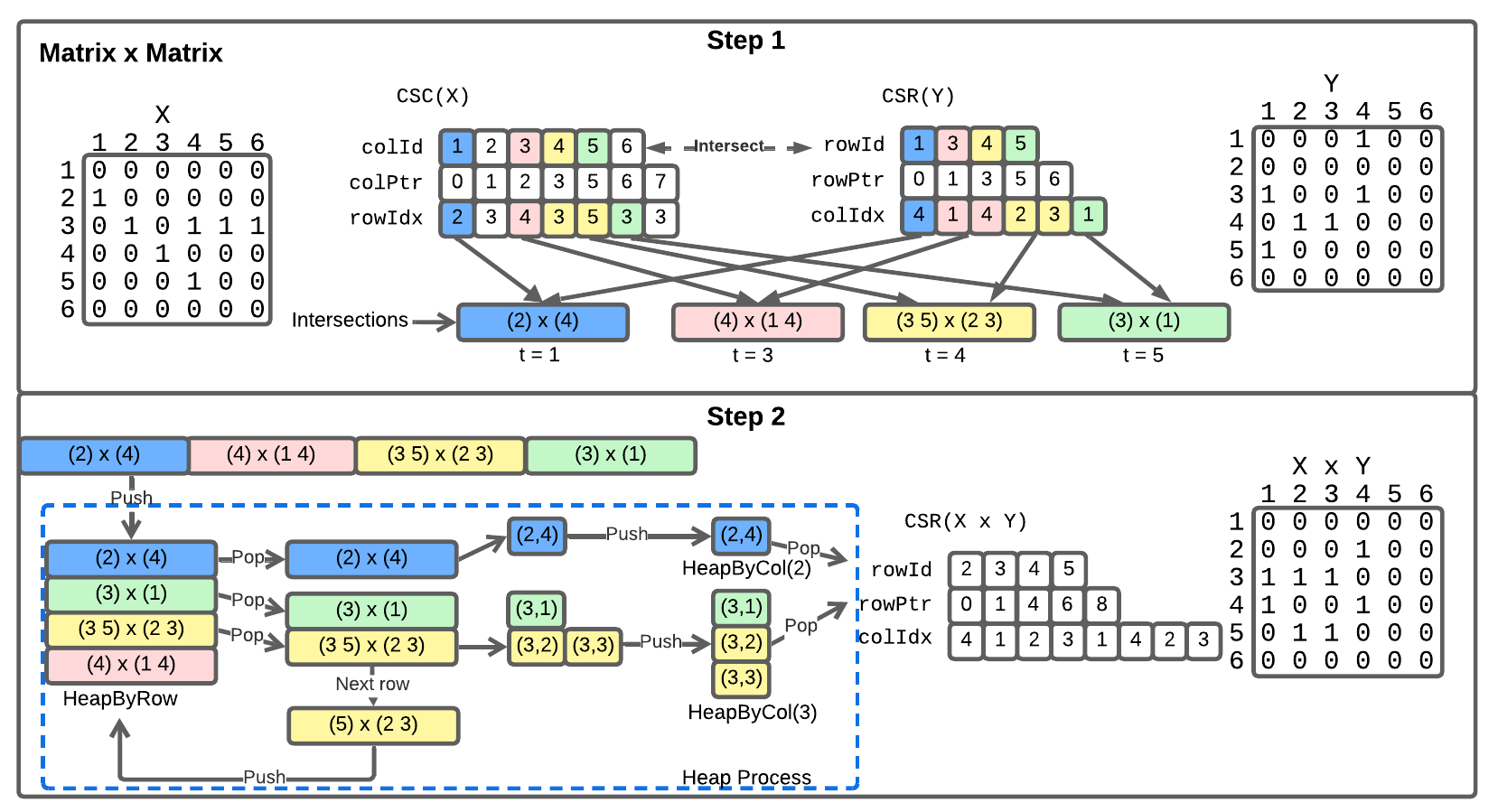}
  \caption{Illustration of Schoor's Boolean matrix multiplication algorithm.}
  \label{fig:steps}
\end{figure}

Figure~\ref{fig:steps} shows an example that illustrates the main stages of the algorithm.
In Step~1, both matrices $X$ and $Y$ are presented in their $\textit{CSC}$ and $\textit{CSR}$ formats, respectively, which are used to compute the multiplication $\textit{CSR}(X \times Y) = \textit{CSC}(X) \times \textit{CSR}(Y)$.
The intersection between the \texttt{colId} array of $\textit{CSC}(X)$ and the \texttt{rowId} array of $\textit{CSR}(Y)$ yields the set $\{1, 3, 4, 5\}$ (highlighted in blue, pink, yellow, and green, respectively); each such element determines a Cartesian product.

For $t = 1$, the algorithm accesses position $0$ in \texttt{colPtr} and retrieves row $\{2\}$ from the \texttt{rowIdx} array; then, from position $0$ in \texttt{rowPtr} it obtains column $\{4\}$.
For $t = 3$, it accesses position $2$ in \texttt{colPtr} and obtains row $\{4\}$; then, from the corresponding position in \texttt{rowPtr} of $\textit{CSR}(Y)$, it retrieves columns $\{1, 4\}$.
For $t = 4$, it retrieves rows $\{3, 5\}$ from \texttt{colPtr} and, from position $3$ in \texttt{rowPtr}, it obtains columns $\{2, 3\}$.
Finally, for $t = 5$, it retrieves row $\{3\}$ from \texttt{colPtr} and column $\{1\}$ from \texttt{rowPtr}.

Step~2 in Figure~\ref{fig:steps} illustrates how Cartesian products are handled using heaps.
Once the intersections are obtained, they are inserted into \texttt{HeapByRow}, which is sorted according to the first element of each row array, ensuring that intersections sharing the same row are grouped together and ordered in ascending form.

For each processed row, a heap \texttt{HeapByCol} is created. In the example, the first row extracted in the first iteration is row $2$.
After performing a \textit{pop} on \texttt{HeapByRow}, the algorithm computes the Cartesian product $(2) \times (4) = (2,4)$, which is inserted into \texttt{HeapByCol(2)}.

In the following two iterations, row $3$ is obtained. Since this row differs from the previous one, \texttt{HeapByCol(2)} is emptied, and as its elements are extracted, the resulting matrix in $\textit{CSR}$ format is constructed.
Next, the Cartesian products $(3) \times (1) = (3,1)$ and $(3\ 5) \times (2\ 3) = (3,2), (3,3)$ are processed and inserted into \texttt{HeapByCol(3)}.
Because the intersection $(3\ 5) \times (2\ 3)$ still has pending rows, it is reinserted into \texttt{HeapByRow} as $(5) \times (2\ 3)$.

In the next iteration, the element extracted from \texttt{HeapByRow} corresponds to $(4) \times (1\ 4)$.
Since this is a new row, \texttt{HeapByCol(3)} is emptied, continuing the construction of the resulting matrix. The final iteration processes $(5) \times (2\ 3)$.
Once both heaps become empty, the algorithm completes the construction of $\textit{CSR}(X \times Y)$.

\section{Current Biclique-based Representation} \label{sec:bicliques}

A representation based on bicliques (i.e., complete bipartite subgraphs) allows a matrix to be decomposed into structured components that enable the implicit representation of edges, reducing the storage and time processing. We describe here the biclique representation we build on \cite{hernandez2014compressed}. 

\begin{definition}
Let \( G = (V, E) \) be a directed graph. The set \( B \) of bicliques in \( G \) is defined as a collection of complete bipartite subgraphs. Each biclique \( b_i(S_i,C_i) \in B \), where \( S_i, C_i \subseteq V \) are subsets of vertices, represents the subgraph \( B_i (S_i \cup C_i, S_i \times C_i) \). The biclique-based representation of graph \( G \) is expressed as the pair \( (B, R) \), where $R = E - \cup_i \,S_i \times C_i$
represents the residual of the graph, that is, that induced by the edges not in any biclique.
\end{definition}

Note that duplicates are allowed, that is, the same edge may appear in more than one biclique $B_i$. This may enable a better representation as a union of bicliques plus a residual graph.

The representation based on bicliques uses $|S_i|+|C_i|$ words to represent the $|S_i| \cdot |C_i|$ edges of biclique $B_i$, and then $2|R|$ words to represent the edges of the residual graph. Those words are written contiguously, biclique by biclique and then the residual graph, in an array. A bitmap is used to mark the first entry of each component $S_i$, $C_i$, and $R$. The total space adds up to $(\sum_i |S_i|+|C_i|)+2|R|$ words, plus a sublinear term for the bitmap. 

The main challenge to use this representation is how to extract suitable bicliques from the graph. Finding the bicliques that yield a minimum-space representation is NP-hard, so heuristics are used. The biclique extraction algorithm used in the current representation~\cite{hernandez2014compressed} aims to extract large bicliques in four steps: Step 1 uses MinHashing~\cite{broder1998min} to compute a small hash set for each adjacency list aiming to efficiently estimate similar adjacency lists. Next, Step 2 builds clusters with similar hash sets, Step 3 sorts by edge frequency each cluster adjacency list to improve the discovery of large bicliques, and Step 4 builds a prefix tree to extract the bicliques.

Their implementation extracts bicliques iteratively using various parameters, in particular the minimum vertex degree to process (\texttt{minAdjNodes}), the number of edges implicitly represented in bicliques to extract (\texttt{bicliqueSize}), a minimum number of bicliques (\texttt{threshold}) below which the extractor reduces the \texttt{bicliqueSize} parameter for the next iteration to enable the extraction of more bicliques, and the maximum number of iterations for the extractor (\texttt{iterations}). This parameterization is done arbitrarily at initialization and does not use information about the input graph; therefore, parameter tuning is cumbersome and impact the quality of the bicliques and the running time required by the algorithm.

\section{Matrix Multiplication with Bicliques}

Boolean matrix multiplication can benefit significantly from the biclique structure of the graphs they represent.
We now present our matrix multiplication algorithm, which adapts Shoor's algorithm (Section~\ref{sec:schoor}) to run on the matrix representation based on bicliques of Section~\ref{sec:bicliques}.

\begin{definition}
Let a graph $G_x$ be represented as a set of bicliques $B$ and a residual graph $R$. We call $B_x$ the matrix that represents the set of bicliques in $B$ and $R_x$ the matrix of its residual graph. The matrix $X$ representing $G$ can then be written as $X = B_x + R_x$.
\end{definition}

Therefore, given graphs $G_x$ and $G_y$, the multiplication $X\times Y$ of their matrices can be computed as
\begin{equation}
 \label{eq:mmult_bicliques}
\begin{aligned}
 X \times Y &= R_x\times R_y + R_x\times B_y + B_x\times R_y + B_x\times B_y.\\
\end{aligned}
 \end{equation}

We will carry out the first multiplication using the algorithm of Arroyuelo et al.~\cite{arroyuelo2025evaluating}; recall Figure~\ref{fig:steps}. The other three components require new algorithms to multiply a general matrix by a set of bicliques, and two sets of bicliques, as we represent those as $b_i(S_i,C_i)$. Those are detailed next.

\subsection{Matrix-to-Biclique Multiplication ($R_x \times B_y$)}
\label{op:axb}
\begin{figure}[t]
  \centering
  \includegraphics[width=1\linewidth]{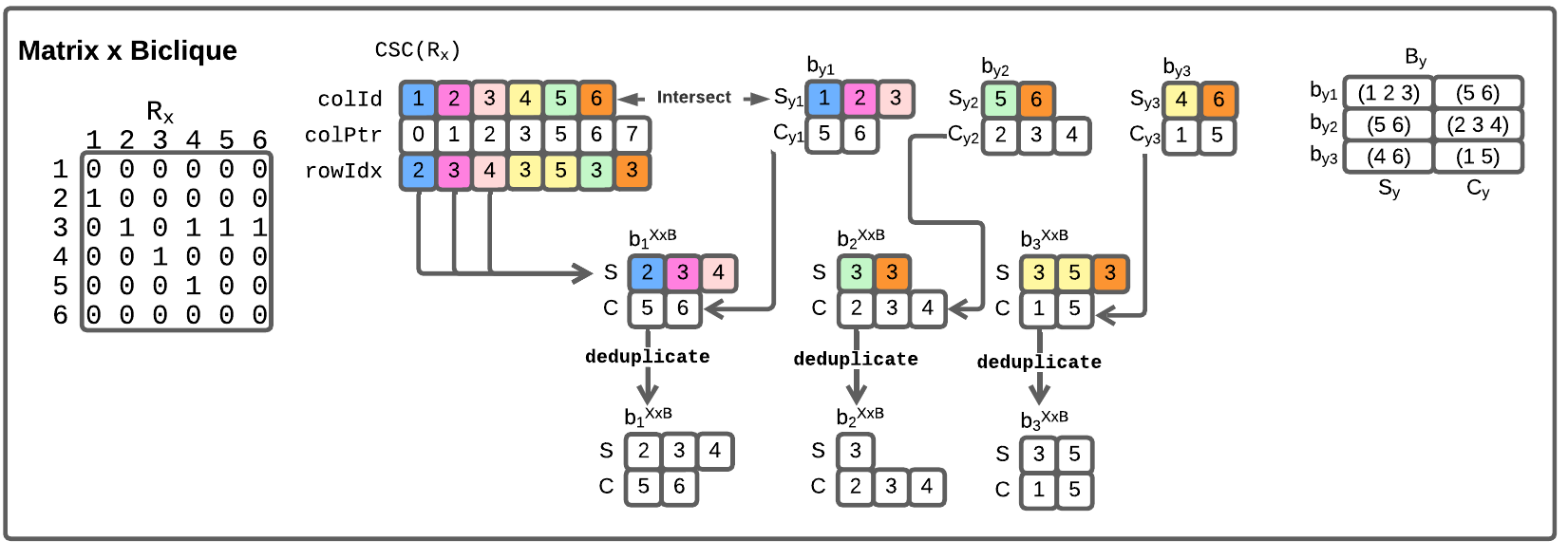}
  \caption{Illustration of the multiplication between $\textit{CSC}(R_x)$ and the set of bicliques $B_y$ in our biclique-based representation. Note edge $(3,5)$ appears twice in the output bicliques.}
  \label{fig:Xxb}
\end{figure}

To perform this multiplication, we compute the product between matrix $R_x$ and the bicliques of matrix $Y$, denoted as $B_y$. For this purpose, we calculate the intersections between the \texttt{colId} array of $\textit{CSC}(R_x)$ and the set $S_{yi}$ of each biclique $b_{yi} \in B_y$. The procedure is as follows; see Figure~\ref{fig:Xxb} for an example.

\begin{enumerate}
    \item For each biclique $b_{yi}$, we identify the intersections between the columns of matrix $R_x$ (in \texttt{colId}) and the rows $S_{yi}$ of the biclique $b_{yi}$.
    \item For each intersecting element $\texttt{colId}[k] \in S_{yi}$, we retrieve the row content from $\texttt{rowIdx}[\texttt{colPtr}[k]..\texttt{colPtr}[k+1]-1]$ of $\textit{CSC}(R_x)$ and the columns in $C_{yi}$ of $b_{yi}$. Note that the product of the rows from $R_x$ and the columns from $C_{yi}$ form indeed a biclique in the output.
    \item Each intersection generates a new biclique, where the rows vary and the columns are always the same set $C_{yi}$. We collapse them all in one single biclique $(S,C_{yi})$, where the sequence $S$ formed by concatenating all the rows from \texttt{rowIdx} of $\textit{CSC}(R_x)$, extracted from the different intersecting elements $\texttt{colId}[k]$.
    \item Finally, the new bicliques undergo a \textit{deduplication} stage on the sequence $S$ to remove repeated rows.
\end{enumerate}

\subsection{Biclique-to-Matrix Multiplication ($B_x \times R_y$)}
\label{op:bxa}

\begin{figure}[t]
  \centering
  \includegraphics[width=1\linewidth]{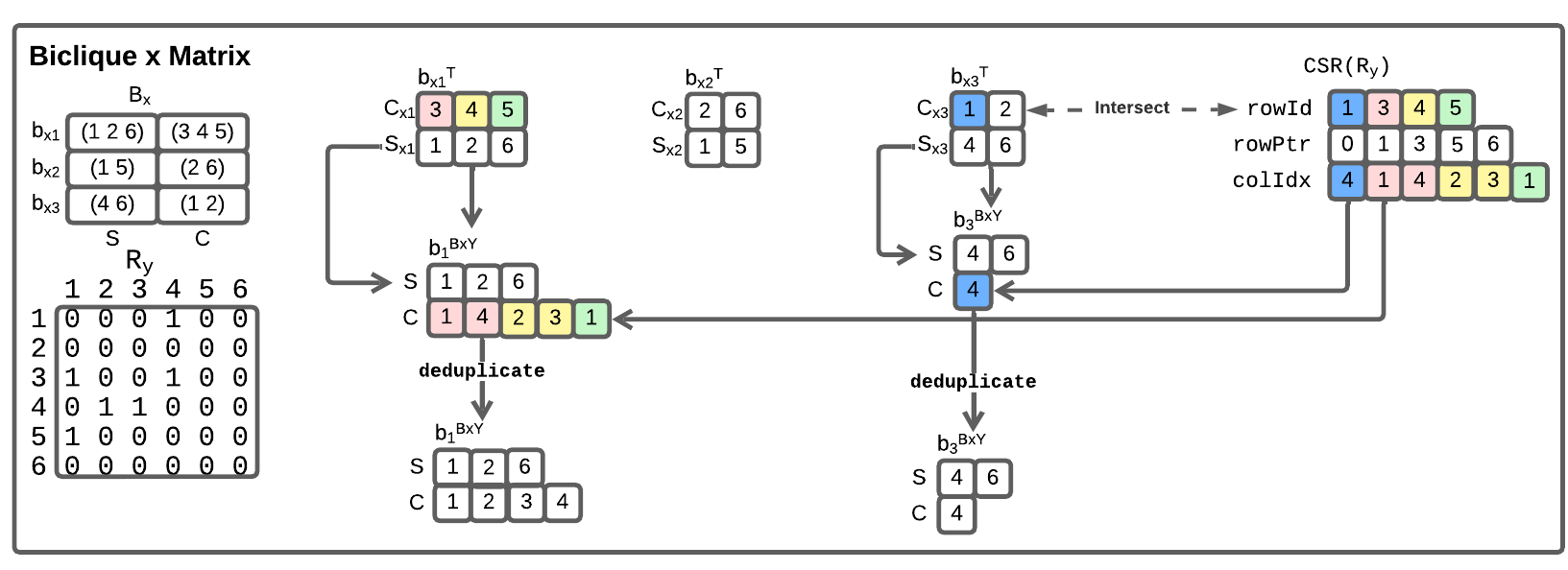}
  \caption{Illustration of the multiplication of the biclique set $B_x$ with the matrix $R_y$.}
  \label{fig:bxa}
\end{figure}

This operation is analogous to the previous one: we now use the $\textit{CSR}$ format for the matrix $R_y$. We then intersect the array $C_{xi}$ of each biclique $b_{xi}$ with the \texttt{rowId} array of $\textit{CSR}(R_y)$ and, for each element in the intersection, form the new bicliques in a manner analogous to the \textit{matrix-to-biclique} case.
To be precise, we go through the following steps; see Figure~\ref{fig:bxa}.

\begin{enumerate}
    \item We compute the intersection between the array $C_{xi}$ of each biclique $b_{xi}$ and the \texttt{rowId} array of $\textit{CSR}(R_y)$.
    
    \item For each intersecting element $\texttt{rowId}[k] \in C_{xi}$, we retrieve the column content from $\texttt{colIdx}[\texttt{rowPtr}[k]..\texttt{rowPtr}[k+1]-1]$ of $\textit{CSR}(R_y)$ and the rows in $S_{xi}$ of $b_{yi}$.
    
    \item The output biclique is $(S_{xi},C)$. To form $C$, we concatenate in a sequence all the qualifying column values from \texttt{colIdx} of $\textit{CSR}(R_y)$.
    
    \item We deduplicate the repeated elements in $C$ for the newly generated bicliques.
\end{enumerate}

\subsection{Biclique-to-Biclique Multiplication}

\label{op:bxb}
\begin{figure}[t]
  \centering
  \includegraphics[width=0.6\linewidth]{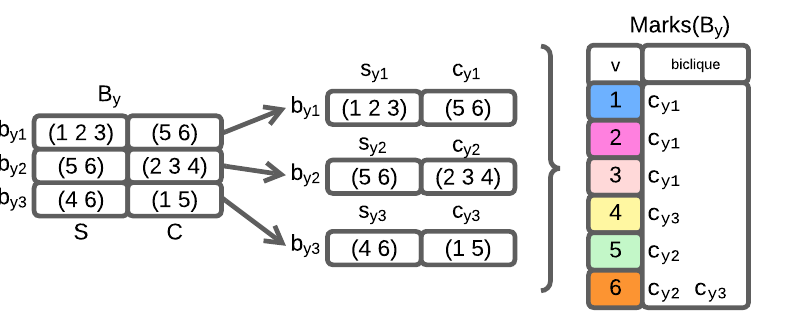}
  \caption{The \textit{marks}$(B_y)$ table, associating each element of any $S_y$ with the sets $C_y$ that connect to it.}
  \label{fig:marksby}
\end{figure}

\begin{figure}[t]
  \centering
  \includegraphics[width=\linewidth]{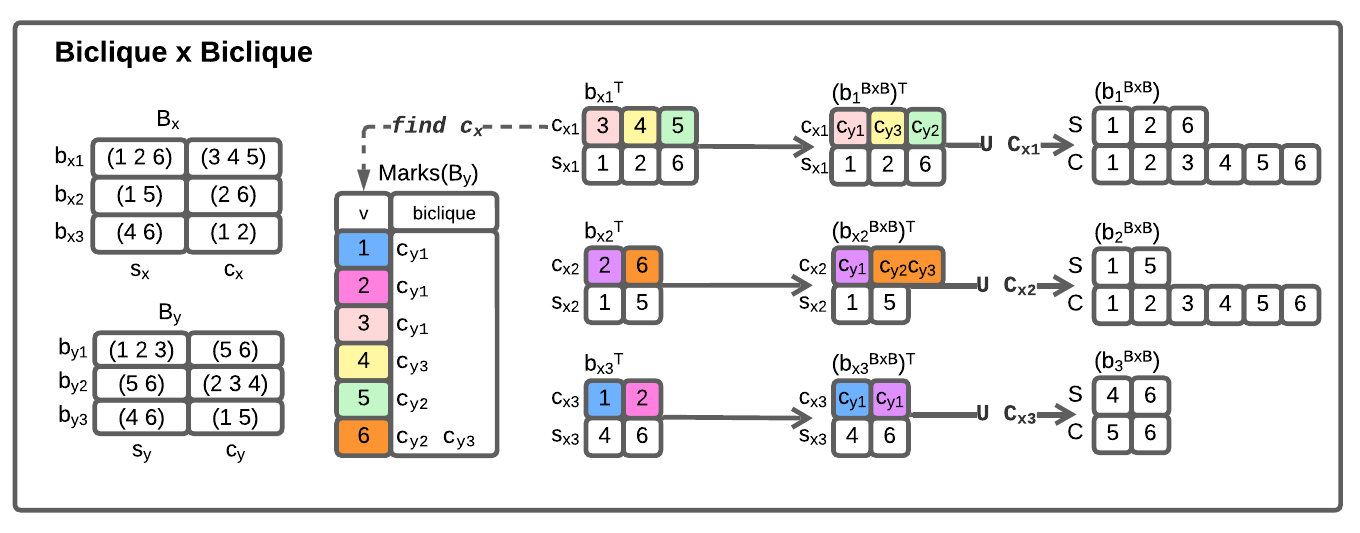}
  \caption{Illustration of the $B_x \times B_y$ multiplication used to propagate adjacency information based on bicliques.}
  \label{fig:bxb}
\end{figure}

To perform the biclique-to-biclique multiplication $B_x \times B_y$, we
first compute the \textit{marks}$(B_y)$ table, shown in Figure~\ref{fig:marksby}, where each element $v$ in any $S_{yi}$ is listed in sorted order and associated with all the sets $C_{yj}$ that connect to it.
The multiplication then proceeds as follows; see Figure~\ref{fig:bxb}.

\begin{enumerate}
  \item For each biclique $b_{xi}$, we search each element of $C_{xi}$ in the \textit{marks}$(B_y)$ table.
  
  \item If the element is found in \textit{marks}$(B_y)$, we obtain its associated content (the corresponding $C_{yj}$ sets). Note that this implies that biclique $(S_{xi},C_{yj})$ belongs to the output.
  
  \item Thus, for every biclique $b_{xi}$ that has at least one intersection, a new biclique $(S_{xi},C)$ is produced: we form the sequence $C$ by combining the $C_{yj}$ values obtained from \textit{marks}$(B_y)$.
  
  \item Finally, duplicated elements in the new set $C$ of each generated biclique are removed.
\end{enumerate}

\subsection{Matrix Input and Output Representation}
\label{representation}

There are various ways in which we can exploit our biclique-based multiplication algorithm. First, we can assume that the input matrices are already compressed with bicliques. Our algorithm produces the output of Eq.~(\ref{eq:mmult_bicliques}) in the form of a residual matrix in {\em CSR} format (for $R_x \times R_y$) and as a set of 
bicliques (for $R_x \times B_y$, $B_x \times R_y$, and $B_x \times B_y$, as shown in this section). We can then represent the matrix $Z = X \times Y$ in the same compressed format:
\begin{eqnarray*}
    B_z &=& R_x \times B_y ~\cup~ B_x \times R_y ~\cup~ B_x \times B_y, \\
    R_z &=& R_x \times R_y.
\end{eqnarray*}
Note that, in this case, $R_z$ may contain some edges that already exist in $B_z$, but the result is still correct. 

In order to reduce space, and because the time to produce $\textit{CSC}(R)$ from $\textit{CSR}(R)$ is very low compared to that of multiplication, we generate and store only $\textit{CSR}(R)$, producing $\textit{CSC}(R)$ on the fly, at multiplication time, when needed. This is in contrast with Arroyuelo et al.~\cite{arroyuelo2025evaluating}, who store both.

Representing the result in the form of bicliques plus a residual matrix has the important property of {\em compositionality}: the output of the multiplication is in the same format as the input, which allows operating it again directly. This enables a matrix algebra where matrices are always represented in this form; it is easy to incorporate other operations like transposition and sums in this format, with boosted efficiency.

\paragraph{Recompression}

One possible drawback of this form of operating is that the biclique representation
$(B_z,R_z)$ may be worse than what we could obtain by extracting bicliques again
from $Z$. For example, bicliques $(S_i,C_i)$ where $|S_i|=1$ or $|C_i|=1$ are better integrated to $R_z$:
after generating
$Z$ in the form above, we can produce $\textit{CSR}(Z)$ from its representation $(B_z,R_z)$, and then find its bicliques as in Section~\ref{sec:bicliques} (in Section~\ref{sec:adaptive} we will show how to do this faster and better).

Recompression can be incorporated to the matrix algebra as an operation that improves the quality of the compression, and applied at any time one deems it convenient. A better representation as bicliques will also speed up further operations on $Z$. In order to apply it, we need to convert the representation from biclique form to \textit{CSR}. This is described in Section~\ref{sec:recovering}.

\paragraph{Bicliques as a speedup}

The speedups for biclique extraction we obtain in Section~\ref{sec:adaptive} enable the use of bicliques not for compression, but just as a device to speed up multiplication. In this case, we assume the matrices are input and output in \textit{CSR} format. We extract bicliques from matrices $X$ and $Y$ before multiplying them, and after multiplication we convert the output to \textit{CSR} form as described in Section~\ref{sec:recovering}. 

\subsection{Recovering the Matrix from the Bicliques} \label{sec:recovering}

\begin{figure}[t]
  \centering
  \includegraphics[width=0.9\linewidth]{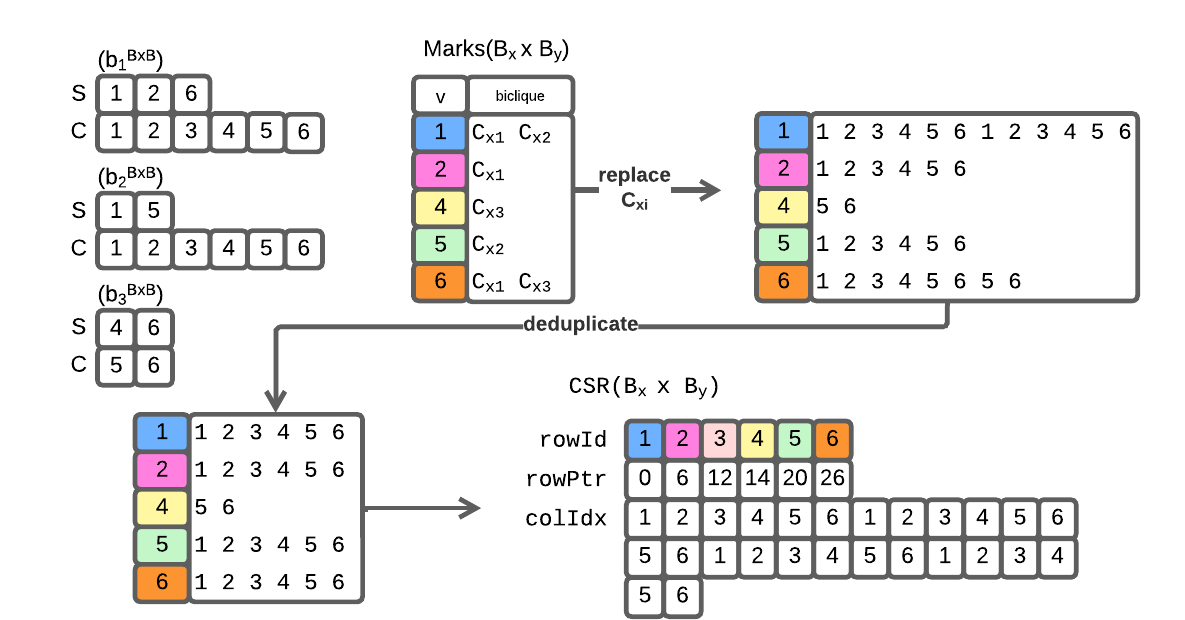}
  \caption{The process of converting the output of $B_x \times B_y$ to \textit{CSR} format.}
  \label{fig:recover}
\end{figure}

To obtain the matrix in \textit{CSR} format from the bicliques computed in the operations described above, we again make use of the \textit{marks} table. From it, we produce the \textit{CSR} representation as follows: we concatenate the elements of the components $C_{zi}$ associated with each row $v$ in \textit{marks}, deduplicate the concatenation, and those are the elements in \texttt{colIdx} corresponding to $v \in \texttt{rowId}$.

Figure~\ref{fig:recover} illustrates the process on the component $B_x \times B_y$; we in fact merge in this way the bicliques generated by $B_x \times R_y$ and $R_x \times B_y$ as well. The result is finally merged (removing duplicates) with the \textit{CSR} representation of the remaining component, $R_x \times R_y$.

\section{Biclique Extraction with Adaptive Parameters} \label{sec:adaptive}

The original biclique extraction technique, described in Section~\ref{sec:bicliques}, uses fixed parameters. 
We now introduce an adaptive parameter setting based on the input graph degree distribution, where some features adapt dynamically between iterations. The goal is to reduce the execution time of the biclique extraction and to improve the edge compression efficiency represented by the extracted bicliques.

To target bicliques with a large number of edges, the idea is to process first nodes with higher degrees, temporarily omitting those with fewer neighbors. This allows the algorithm to reduce the computation time in all its steps. We found that an effective way to define this pruning is by applying percentile-based filtering over the degree distribution. For instance, setting the minimum degree threshold at the 90th percentile restricts the search to the top 10\% of nodes with the highest connectivity.

\begin{figure}[t!]
  \centering
  \includegraphics[width=1\linewidth]{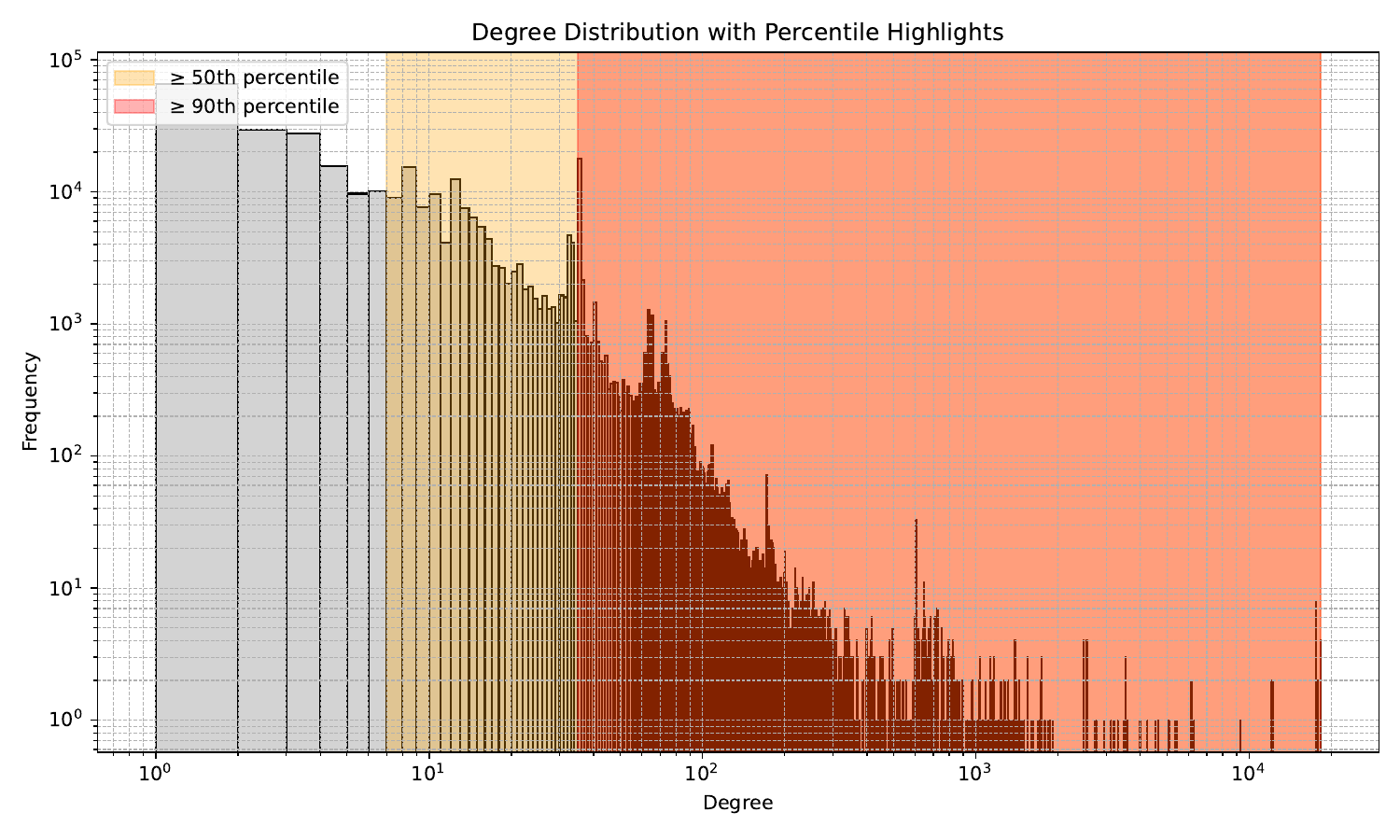}
  \caption{Out-degree distribution of the \textit{cnr-2000-hc} graph grouped by percentiles, emphasizing dense regions suitable for biclique extraction.}
  \label{fig:cnrdistperc}
\end{figure}

As an example, Figure~\ref{fig:cnrdistperc} shows the degree distribution of the \textit{cnr-2000-hc} graph~\cite{boldi2004webgraph, BRSLLP},
segmented by percentiles. The yellow region highlights nodes within the 50th percentile, while the red region marks those in the 90th percentile. This identifies the ranges of nodes that are most relevant for extracting largest bicliques, helping guide the initial parameterization of the algorithm.

Algorithm~\ref{opt_alg} presents the pseudocode for the iterative biclique extraction process, using percentiles as an adaptive tuning mechanism. The procedure begins with an initial traversal of the graph to build a degree distribution, which helps identify highly connected nodes.
Based on this distribution, the parameter \texttt{minAdjNodes}, which filters out the small degree adjacency lists, is estimated using the 90th percentile. This initial filter favors the detection of large bicliques.

\begin{algorithm}[t!]
\scriptsize
%\label{opt_alg}
\begin{algorithmic}[1]
\Require Graph \( G \)
\Ensure Residual graph \( G' \), set of bicliques \( B_G \)

\State Initialize: \texttt{unordered\_map<int, int> distribution}

\Comment{Step 1: Build adjacency distribution}
\ForAll{node \( v \in G \)}
    \State \( \texttt{distribution[gradeOf}(v)\texttt{]} \gets \texttt{distribution[gradeOf}(v)\texttt{]} + 1 \)
\EndFor

\Comment{Step 2: Initialize parameters}
\State \( \texttt{minAdjNodes} \gets \texttt{getPercentile}(90) \)
\State \( \texttt{currIt} \gets 0 \)
\State \( \texttt{totalBicliques} \gets 0 \)
\State \( \texttt{threshold} \gets 0.8 \)

\Comment{Step 3: Iterative extraction}
\While{ \( \texttt{currIt} < \texttt{maxIt} \) }
    \State \( \texttt{signatures} \gets \texttt{computeMinHashes(minAdjNodes)} \)
    \State \( \texttt{clusters} \gets \texttt{clusterize(signatures)} \)
    \State \( \texttt{bicliqueExtracted} \gets \texttt{extractBicliques(clusters)} \)\\
    \State \( B_G \gets B_G + \texttt{bicliquesExtracted} \)
    \State \( \texttt{avgBicliques} \gets \texttt{B\_G.size()} / \texttt{currIt} \) 
    
    \Comment{Step 4: Decide new parameters}
    \If{ \( \texttt{bicliquesExtracted.size()} < \texttt{avgBicliques} \cdot \texttt{threshold} \) }
        \State \( \texttt{minAdjNodes} \gets \texttt{minAdjNodes} \cdot \texttt{threshold} \)
    \EndIf
    \State \( \texttt{currIt} \gets \texttt{currIt} + 1 \)
\EndWhile
\end{algorithmic}
\caption{Adaptive Biclique Extraction Process. Each step adjusts parameters based on intermediate statistics to balance extraction time and compression effectiveness.}
\label{opt_alg}
\end{algorithm}

The algorithm then proceeds with the iterative extraction phase, following the original approach~\cite{hernandez2014compressed}. At the end of each iteration, the number of extracted bicliques is compared with the historical average, adjusted by a tolerance threshold (\texttt{threshold} = 0.8). If there is a reduced productivity, the \texttt{minAdjNodes} parameter is reduced by 20\%, allowing the algorithm to explore nodes with lower degrees in subsequent rounds.

This approach enables a segmented search for bicliques, which reduces the computational cost of extraction by processing fewer elements per iteration. Moreover, the selected nodes tend to share similar degrees, enhancing the structural coherence of the extracted bicliques.

\section{Experimental Evaluation}

\begin{table}[t]
\centering \small
\begin{tabular}{|l|r|r|r|}
\hline
\textbf{Dataset} & \textbf{$|V|$} & \textbf{$|E|$} & \textbf{$|E|/|V|$} \\
\hline
web-Stanford         & 281,903     & 2,312,497     & 8.20    \\
cnr-2000-hc          & 325,557     & 3,126,152     & 9.87    \\
wikipedia-link-lmo & 52,214      & 3,623,678     & 64.40   \\
web-Google           & 875,713     & 5,105,039     & 5.83    \\
coPapersCiteseer     & 434,102     & 32,073,436    & 73.88    \\
indochina-2011       & 7,414,866    & 194,109,311   & 26.17   \\
arabic-2005          & 22,744,080   & 639,999,458   & 28.13   \\
\hline
\end{tabular}
%}
\caption{Main statistics of the graphs used in the experiments, including number of nodes, number of edges, and average degree.}
\label{tab:datasets}
\end{table}

We built on the \texttt{C++} implementation of Arroyuelo et al.~\cite{arroyuelo2025evaluating}, extending the algorithm to support biclique operations.
We measured time and space usage of biclique compression and execution time of multiplications, comparing our results under various scenarios with the version of Schoor's algorithm implemented by Arroyuelo et al., which we call \textit{Baseline}. We also include a comparison with the $k^2$-tree compressed representation by the same authors.

The datasets used are publicly available from various sources, such as The Laboratory of Web Algorithmics (LAW), collected and used by Boldi and Vigna~\cite{boldi2004webgraph}, the Stanford Network Analysis Platform (SNAP)~\cite{snapnets}, and the Network Repository~\cite{nr}. Table~\ref{tab:datasets} presents the main statistics for the chosen datasets, including the number of vertices $|V|$, the number of edges $|E|$, and the node degree average, $|E|/|V|$.

All results were obtained by using sequential implementations, running on an \textit{AMD EPYC 9354} Linux server with 256 GB of main memory.

The source code of our approach is available in the Github repository \url{https://github.com/nicolasaraya/BicliqueMultiplication}.

\subsection{Biclique Extraction and Compression}

We first compare the space compression and running time performance between the adaptive (Section~\ref{sec:adaptive}) and non-adaptive (the original implementation \cite{hernandez2014compressed}, see Section~\ref{sec:bicliques}) versions of the biclique extractor algorithm. 
To perform the comparison, we define the compression ratio as the percentage of the space saved by compression (higher is better):
$$Ratio = 100\left(1 - \frac{(\sum_i |S_i| + |C_i|) + |R|}{|E|}\right).$$

\begin{table}[tb]
\centering \small
%\resizebox{\textwidth}{!}{%
\begin{tabular}{|l|r|r|r|r|r|r|}
\hline
\multirow{2}{*}{\textbf{Dataset}} &  \multicolumn{2}{c|}{\textbf{Ratio (\%)}} & \multicolumn{2}{c|}{\textbf{BicEx (s)}} & \multicolumn{2}{c|}{\textbf{bpe(B+R)}} \\
\cline{2-7}
 & Non & Adaptive & Non & Adaptive & Non & Adaptive \\
\hline
web-Stanford           
& 21.64 & \textbf{52.35} & 2.54 & \textbf{2.40} & 32.46 & \textbf{23.21} \\
cnr-2000-hc           
& 41.32 & \textbf{59.60} & 2.71 & \textbf{1.17} & 23.61 & \textbf{17.88} \\
wikipedia-link-lmo   
& \textbf{67.15} & 65.95 & 2.95 & \textbf{1.35} & \textbf{11.43} & 11.83 \\
web-Google             
& 2.01 & \textbf{40.98} & \textbf{8.40} & 13.58 & 40.59 & \textbf{28.88} \\
coPapersCiteseer       
& 49.94 & \textbf{64.47} & 47.38 & \textbf{13.69} & 16.89 & \textbf{12.29} \\
indochina-2011         
& 73.96 & \textbf{78.33} & 173.12 & \textbf{89.73} & 10.18 & \textbf{8.85} \\
arabic-2005-hc         
& 70.75 & \textbf{74.13} & 951.66 & \textbf{441.95} & 11.17 & \textbf{10.18} \\
\hline
\end{tabular}
%}
\caption{Comparison between biclique extraction with our adaptive and the previous non-adaptive parameter configurations. Column \textit{BicEx} gives the running time in seconds of the biclique extraction process. Column bpe gives the bits per edge of the resulting representations. The best values are highlighted in bold.}
\label{tab:extraction_opt}
\end{table}

\begin{figure}[tb]
  \centering
  \includegraphics[width=1\linewidth]
  {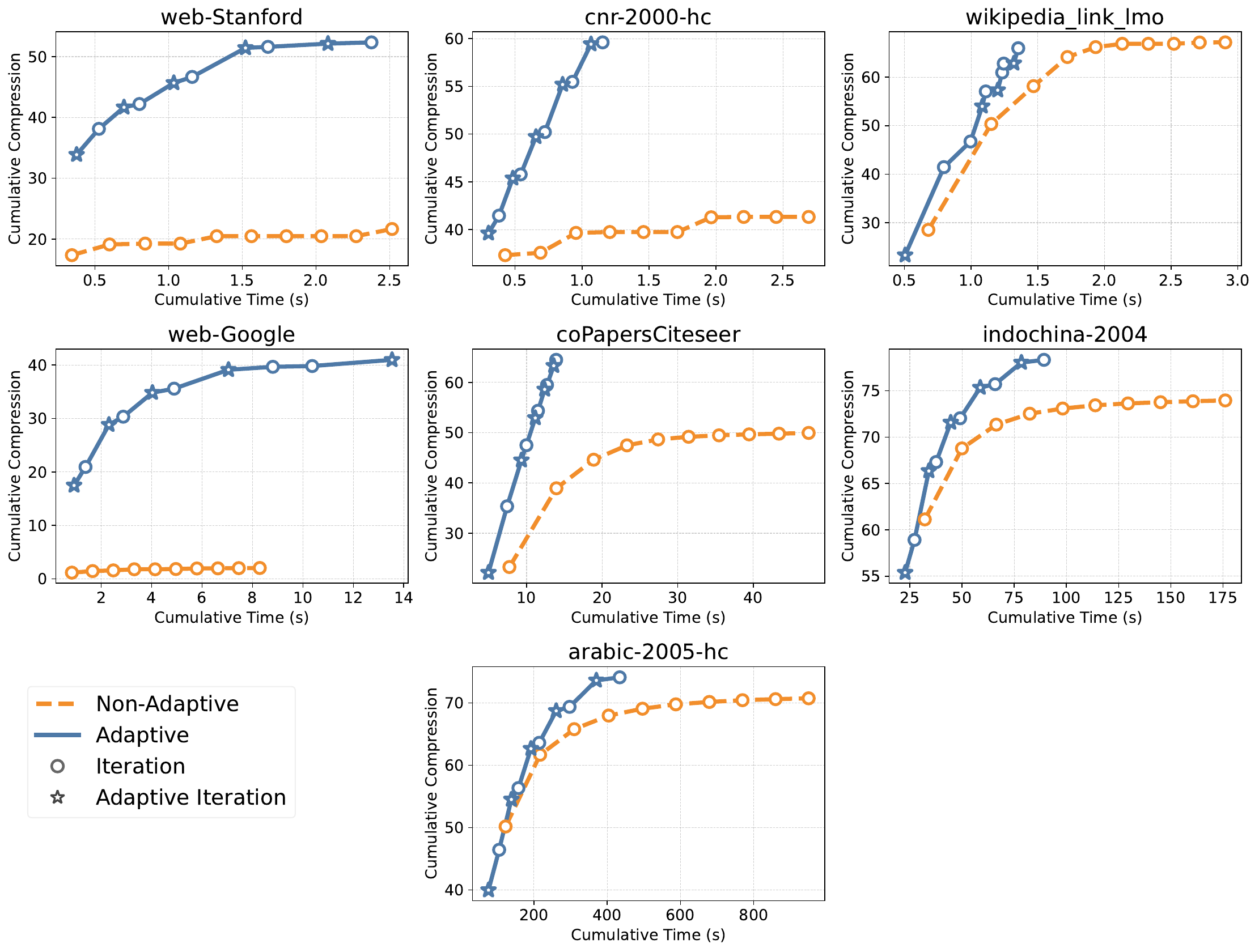}
    \caption{Cumulative compression per second for adaptive and non-adaptive biclique extraction on each dataset.}
  \label{fig:subPlot}
\end{figure}

Table~\ref{tab:extraction_opt} compares the compression and execution time performances obtained using our adaptive biclique extraction method against the non-adaptive alternative using fixed parameters and ten iterations. We observe that adaptive outperforms non-adaptive extraction in both compression ratio and speed in almost all cases. It usually takes 45\% of the time of the non-adaptive version (and as little as 29\%), being slower in only one case, where it takes 162\% of the time. In that case, however, the non-adaptive version is faster because it fails to compress, obtaining a compression ratio of just 2\%, versus 41\% of the adaptive version. The resulting compressed representation usually takes 70\%-75\% of the space required by the one obtained with the non-adaptive version, being larger in only one case, where it takes 103\% of the space.  

Figure~\ref{fig:subPlot} shows how compression ratios grow over time, along the successive iterations, in both adaptive and non-adaptive implementations. In all cases, the adaptive version achieves a higher cumulative compression ratio per second, indicating that dynamically adjusting the extraction parameters accelerates convergence while maintaining superior compression efficiency. The adaptive case shows the first iteration after each parameter adaptation using a star symbol; it can be seen that such iteration effectively improves compression without significantly worsening the execution times in most datasets.

\begin{table}[tb]
\centering 
\resizebox{\textwidth}{!}{%
\begin{tabular}{|l|r|r|r|r|r|}
\hline
\textbf{Dataset} & \textbf{bpe($G_M$)} & \textbf{bpe($B+R$)} & \textbf{bpe($G_{k2}$)} & \textbf{bpe($G_{RP32}$)} & \textbf{bpe($G_{WG}$)}\\
\hline
\footnotesize{web-Stanford} &  76.68 & 23.21  & 26.37 & 20.00 & 19.74 \\
\footnotesize{cnr-2000-hc}  &  73.62 & 17.88  & 3.49 & 13.46 & 2.89   \\
\footnotesize{wikipedia-link-lmo} & 65.41 & 11.83  & 2.87 & 3.64 & 1.42    \\
\footnotesize{web-Google} & 79.66 & 28.88   & 33.57 & 24.66 & 23.88   \\
\footnotesize{coPapersCiteseer} &  65.54 & 12.29  &  3.24 & 4.24 & 1.63\\
\footnotesize{indochina-2011} & 68.17 &  8.85   &2.40 & 5.52 & 1.45  \\
\footnotesize{arabic-2005} & 68.08 &  10.18   & 2.74 & 5.72 & 1.84\\
\hline
\end{tabular}
}
\caption{Bits per edge comparison between different compression methods: bpe($G_M$) corresponds to the baseline representation, bpe$(B+R)$ to our technique, bpe($G_{k2}$) to the $k^2$-tree, bpe($G_{RP32}$) to RePair32, and bpe($G_{WG}$) to WebGraph.}
\label{tab:bpes}
\end{table}

Next, we compare the compression efficiency of our biclique representation (using adaptive construction) with other compression techniques. Table~\ref{tab:bpes} shows the bpe (bits per edge) for the biclique-based representation, bpe$(B+R)$, and compares it with bpe$(G_M)$, the baseline graph representation used for matrix multiplication by Arroyuelo et al.~\cite{arroyuelo2025evaluating}, as well as their $k^2$-tree implementation, bpe$(G_{k^2})$.
We also include bpe$(G_{RP32})$, the compression obtained by RePair32~\cite{ferragina2022improving}, and bpe$(G_{WG})$, the Webgraph framework ~\cite{boldi2004webgraph}.

As observed, Webgraph achieves the best compression, followed by $k^2$-tree and RePair32. Those outperform our technique in space, by a large margin in many cases. Among the representations for which matrix multiplication is supported directly, however, our biclique compression substantially reduces the space of the baseline across all datasets, using 13\%--36\% of its space (even if we improved the space of the baseline by storing only the \textit{CSR} part, we would use 26\%--72\% of its space). The $k^2$-tree also supports matrix multiplication, but as we will see, it can be orders of magnitude slower than our biclique-based representation.

\subsection{Matrix Multiplication using Bicliques only for Operations }

This section studies the usage scenario where the matrices are still stored in \textit{CSR} format, which is then the input and output format of the multiplication algorithm. Our biclique representation is used exclusively to speed up multiplications, not to compress. Therefore, finding the bicliques is part of the multiplication process (and so, the extraction cost is accounted for). 

Because matrices in our dataset are of different sizes, we will measure the time to compute $X^2$ on individual matrices $X$ in order to evaluate performance of multiplications of the form $X\times Y$. We measure all execution times as if we multiplied two different matrices, for example we do not exploit the fact that we only need to extract bicliques from one matrix.  

Table~\ref{tab:scenario1} presents the execution time of our method with the different steps required to perform the multiplication. Column \textit{BicEx} shows the execution time of the biclique extraction for the two matrices, which in practice we measured as counting twice the execution time on matrix $X$. Next, we give the time required to build the \textit{CSC} format from the \textit{CSR} of the residual matrix $R_x$, needed to compute $R_x\times R_y$. Then, we show the multiplication time used for each of the four components, the time required to merge the results into a biclique-based representation, and finally the time to convert that to \textit{CSR} form. Column \textit{Total} adds up all these times and can be compared to the final column, which shows the multiplication time with the \textit{Baseline}.

\begin{table}[tb]
\centering
\resizebox{\textwidth}{!}{%
\begin{tabular}{|l|r|r|r|r|r|r|r|r|r|r|}
\hline
\multirow{2}{*}{\textbf{Dataset}} & \multicolumn{9}{c|}{\textbf{Our approach (s)}}  & \multirow{2}{*}{\textbf{Baseline (s)}} \\
\cline{2-10}
& BicEx & CSC$(R_x)$ & $R_x\times R_y$ & $R_x\times B_y$ & $B_x\times R_y$ & $B_x \times B_y$
& Merge & to CSR & Total &  \\
\hline
web-Stanford       &
4.80 & 0.04 & 0.16 & 0.03 & 0.04 & 0.17 & 0.47 & 0.85 & 6.56 & 1.27\\
cnr-2000-hc        &
2.34 & 0.03 & 0.19 & 0.01 & 0.03 & 0.09 & 0.12 & 0.76 & 3.57 & 1.34\\
wikipedia-link-lmo &
2.70 & 0.02 & 0.66 & 0.02 & 0.05 & 0.20 & 0.01 & 0.46 & 4.12 & 9.52\\
web-Google         &
27.16 & 0.22 & 0.58 & 0.14 & 0.18 & 0.56 & 1.83 & 2.28 & 32.95 & 2.43\\
coPapersCiteseer   &
27.38 & 0.19 & 4.39 & 2.35 & 2.50 & 12.13 & 2.85 & 7.09 & 58.88 & 91.99\\
indochina-2011     &
179.46 & 1.29 & 8.68 & 0.78 & 1.62 & 11.34 & 7.33 & 51.35 & 261.85 & 8982.77\\
arabic-2005        &
883.90 & 4.89 & 44.29 & 3.53 & 7.55 & 45.43 & 36.62 & 220.08 & 1246.29  & 2175.63\\
\hline
\end{tabular}
}
\caption{Matrix multiplication using only bicliques while preserving the \textit{CSR} representation. Times are reported in seconds. The total execution time includes biclique extraction (once per matrix), a \textit{CSC} conversion for matrix $R_x$, the four multiplication operations, the merge step to obtain the $(B + R)$ representation, and a final step to convert the result into \textit{CSR} format. The execution time of the Baseline is also reported.}
\label{tab:scenario1}
\end{table}

The results show that our approach can be much faster than the \textit{Baseline}, achieving speedups between and $1.75$ and $34.3$ on the larger datasets, while the \textit{Baseline} is typically much faster on the smaller datasets. Our technique then has a clear impact on the hardest multiplications, which take minutes to hours, while it loses, less relevantly in absolute terms, on easy multiplications that take seconds. Our improvements on the performance of biclique extraction is key to make this process worthy, as its cost is compensated by a much faster multiplication.

\subsection{Matrix Multiplication using the Biclique Representation}

\begin{table}[tb!]
\centering
\resizebox{\textwidth}{!}{%
\begin{tabular}{|l|r|r|r|r|r|r|r|r|}
\hline
\multirow{2}{*}{\textbf{Dataset}} & \multicolumn{2}{c|}{\textbf{Baseline}} & \multicolumn{2}{c|}{\textbf{k$^2$-tree}}
& \multicolumn{4}{c|}{\textbf{Our approach}}\\
\cline{2-9}
& bpe & time(s) & bpe & time(s) & bpe & Mult(s) & BicEx2(s) & Total(s) \\
\hline
web-Stanford    &
65.47 & 1.27 & 20.45 & 283.44 & 7.92 & 1.76 & 6.05 & 7.81 \\
cnr-2000-hc     &
64.96 & 1.34 & 2.61 & 6.63 & 5.86 & 1.23 & 6.60 & 7.83 \\
wikipedia-link-lmo &
64.26 & 9.52 & 3.73 & 24.30 & 9.73 & 1.42 & 7.91 & 9.33 \\
web-Google      & 
66.63 & 2.43 & 27.99 & 1199.00 & 12.47 & 5.79 & 14.49 & 20.28 \\
coPapersCiteseer& 
64.19 & 91.99 & 3.17 & 324.62 & 8.31 & 31.50 & 113.85 & 145.35 \\
indochina-2011  &
64.43 & 8982.77 & 2.47 & 4519.20 & 5.74 & 82.39 & 488.15 & 570.54 \\
arabic-2005     &
64.32 & 2175.64 & 2.68 & 3579.26 & 6.16 & 362.39 & 2435.97 & 2798.36 \\
\hline
\end{tabular}
}
\caption{Space in bpe and multiplication time in seconds using our method with the $R+B$ format for both input and output, with recompression. Column \textit{Mult} reports the total time to multiply the biclique-based representations and produce the result in \textit{CSR} form, as presented in Table~\ref{tab:scenario1}.
Column \textit{BicEx2} gives the additional time to apply biclique extraction on the result, and column \textit{Total} gives the total time. For comparison purposes, we also include the bpe and multiplication times of the \textit{Baseline} and the \textit{$k^2$-tree}.}
\label{tab:scenario2}
\end{table}

We now consider the second usage scenario, where bicliques are used to represent matrices using less space, and therefore the multiplication algorithm uses our biclique-based representation for both input and output. We do not then carry out biclique extraction on the input matrices, as they already come in this format, but we  {\em recompress} the output matrix, finding its bicliques from the \textit{CSR} output. We compare the resulting space and execution time of our approach with those of the \textit{Baseline} and the $k^2$-tree.   

Table~\ref{tab:scenario2} shows that our approach with recompression provides a great compression improvement on the output compared to the \textit{Baseline}, using 9\%--19\% of the space (and 18\%--38\% if we improved the {\em Baseline} to store only the \textit{CSR} part). On the larger matrices, where time performance matters most, our technique is competitive with the {\em Baseline}, being only 30\%--60\% slower (and 16 times faster in one case). On the smaller matrices, we are slower by a factor of up to 8. Our technique is, instead, almost always faster than the $k^2$-tree, by up to 59 times; the $k^2$-tree is faster only on one small matrix, by 18\%. On the other hand, the $k^2$-tree output is generally smaller than our representation by a factor of 2.3--2.6, though the situation is reversed in a couple of datasets.

\begin{table}[tb!]
\centering \small
\begin{tabular}{|l|r|r|r|}
\hline
\textbf{Dataset} & $E(X)$ & $E(X^2)$ & $E(X^4)$ \\
\hline
web-Stanford & 2,312,497 & 20,811,442 & 429,584,958 \\
cnr-2000-hc & 3,216,152 & 34,174,066 & 683,240,962 \\
wikipedia-link-lmo & 3,623,678 & 19,997,655 & 549,950,901 \\
web-Google & 5,105,039 & 29,710,164 & 585,573,462 \\
coPapersCiteseer & 32,073,436 & 264,584,716 & 12,248,543,674 \\
indochina-2011 & 194,109,311 & 1,952,630,542 & - \\
arabic-2005 & 639,999,458 & 8,323,612,632 &  - \\
\hline
\end{tabular}
\caption{Number of edges for the graph representing $X$, $X^2$, and $X^4$. }
\label{tab:x_edges}
\end{table}

Comparing columns {\em Mult} and {\em BicEx2}, we note that the need to recompress the output makes our biclique-based multiplication considerably slower. Table~\ref{tab:x_edges} explains why it is much more costly to extract bicliques on the output than it was on the input matrices: they tend to be considerably larger in our case (i.e., proportional to the number of paths of length 2 in the represented graph). This motivates a third scenario, where we take the bicliques output by the multiplication algorithm directly as the output representation, without recompression.

\subsection{Matrix Multiplication with Bicliques and no Recompression}

\begin{table}[tb!]
\centering
\small
\begin{tabular}{|l|r|r|r|r|r|r|}
\hline
\multirow{2}{*}{\textbf{Dataset}} &
\multicolumn{2}{c|}{\textbf{Baseline}} & 
\multicolumn{2}{c|}{\textbf{Recompression}} & \multicolumn{2}{c|}{\textbf{No Recompression}} \\
\cline{2-7}
& bpe & time(s) & bpe & time(s) & bpe & time(s) \\
\hline
web-Stanford    &
65.47 & 1.27 & 7.92 & 7.81 & 11.03 & 0.91\\
cnr-2000-hc     &
64.96 & 1.34 & 5.86 & 7.83 & 8.22 & 0.47  \\
wikipedia-link-lmo &
64.26 & 9.52 & 9.73 & 9.33 & 11.74 & 0.96  \\
web-Google      & 
66.63 & 2.43 & 12.47 & 20.28 & 17.06 & 3.51 \\
coPapersCiteseer& 
64.19 & 91.99 & 8.31 & 145.35 & 7.99 & 24.41 \\
indochina-2011  &
64.43 & 8982.77 & 5.74 & 570.54 & 6.15 & 31.04  \\
arabic-2005     &
64.32 & 2175.64 & 6.16 & 2798.36 & 7.47 &  142.31 \\
\hline
\end{tabular}
%}
\caption{Comparison of space (in bpe) and multiplication time (in seconds) with and without recompression. Column \textit{Recompression} reports the space and execution time of our method when multiplication is followed by biclique extraction on the output. Column \textit{No Recompression} shows the performance when the result is directly left in the $B+R$ format produced by our biclique-based multiplication algorithm, without recompressing.}

\label{tab:scenario3}
\end{table}

We finally evaluate the usage scenario where we directly use the bicliques output by the multiplication as the biclique-based representation of the output. This yields a much faster multiplication algorithm, as bicliques are never extracted. Instead, a new issue appears: the bicliques may degrade (i.e., more and smaller bicliques, a larger residual matrix) when applying a multiplication, compared to those obtained by recompression. We measure those aspects. We note that, still, we are free to apply recompression at any time when the quality of some representation is deemed to have degraded.

Table~\ref{tab:scenario3} shows the space and execution time comparing the multiplication with and without recompression. It can be observed that the compressed space degrades in most cases by 7\%-40\% when recompression is not applied (curiously, recompression slightly worsens the space in one case). In exchange, multiplication is much faster with no recompression, by a factor of up to 20. Compared to the {\em Baseline}, this alternative uses only 10\%--26\% of the space (or 20\%--52\% if we improved the {\em Baseline} space) and is almost always faster, by a factor of up to 289 (it is slower only in one case, by 44\%).

Our final experiment aims to measure how the degradation in the quality of the bicliques may affect the time performance of future operations on the matrix. For this sake, we square matrices $X$ and then square the resulting $X^2$, to obtain $X^4$. We do this with and without an intermediate recompression. In the first case, we recompress $X^2$ before squaring it again, with the hope that this effort is compensated by the biclique-based multiplication finding bigger bicliques.

\begin{table}[tb]
\centering
\resizebox{\textwidth}{!}{%
\begin{tabular}{|l|rr|rr|rrrr|rrrrr|}
\hline
\multirow{2}{*}{\textbf{Dataset}} & \multicolumn{2}{c|}{\textbf{Baseline (s)}} & \multicolumn{2}{c|}{\textbf{k$^2$-tree (s)}} & \multicolumn{4}{c|}{\textbf{No recompression (s)}} & \multicolumn{5}{c|}{\textbf{Recompression (s)}} \\
\cline{2-14}
& $X^4$ & bpe &
$X^4$ & bpe &
$X^2$ & $X^4$ & Total  & bpe &
$X^2$ & BicEx2 & $X^4$ & Total & bpe \\
\hline
web-Stanford & 
81.78 & 64.07 &
8,898.64 & 12.65 &
0.91 & 22.74 & 23.65   & 9.00 & %no recom
1.76 & 6.05 & 7.38 & \textbf{15.19} & 4.93 \\
cnr-2000-hc & 
198.85 & 64.05 &
455.96  & 1.70 &
0.47 & 28.96 & 29.43  & 5.82 & %no recom
1.23 & 6.60 & 6.75 & \textbf{14.58} & 3.56\\
wikipedia-link-lmo & 
335.09 & 64.01 &
1,008.50  & 2.41 &
0.96 & 79.49 & 80.45  & 16.45 & % no recom
1.42 & 7.91 & 45.00 & \textbf{54.33} & 14.57 \\
web-Google & 
87.32 & 64.14 &
19,879.91 & 19.14 &
3.51 & 52.88 & 56.39 & 15.14 & % no recom
5.79 & 14.49 & 16.79 & \textbf{37.07} & 8.05 \\
coPapersCiteseer &
13,398.54 & 64.00 & 
26,775.85 & 2.40 &
24.41 & 2,948.08 & 2,972.49 & 7.23 & %no recom
31.50 & 113.85 & 1,039.51 & \textbf{1,184.86} & 6.63\\
\hline
\end{tabular}%
}
\caption{Comparison of execution times (in seconds) and space for computing $X^4$ with and without intermediate recompression. }
\label{tab:x4_comparison}
\end{table}

Table~\ref{tab:x4_comparison} presents the space and execution times with both strategies. We also include the \textit{Baseline} and the $k^2$-tree. 
Column \textit{No recompression} includes the two multiplication operations and the total. Column \textit{Recompression} also includes the recompression for $X^2$ (but not for $X^4$). As observed, the two alternatives of our approach are faster than using the \textit{Baseline} (by a factor up to 13.6) and $k^2$-tree (by a factor up to 586). On the other hand, recompressing $X^2$ makes the whole process almost twice as fast as not recompressing it, that is, the intermediate recompression pays off. In terms of compression efficiency, our approach is many orders of magnitude more efficient than the \textit{Baseline} in all matrices (by a factor over 55 in a couple of matrices), and even smaller than the $k^2$-tree in a couple of them. 

The outcome of this experiment is better understood when considering that matrices tend to become considerably denser when multiplied.
Table~\ref{tab:x_edges} shows the total number of edges of the input matrix $X$, in $X^2$, and in $X^4$. It can be seen that the number of edges increases by an order of magnitude each time we square the matrix. The table omits the two largest matrices, as we could not compute $X^4$ due to memory limitations. Further, the fact that we outperformed the {\em Baseline} in space by a much wider margin suggests that matrices that are products of others are much more compressible via bicliques, which is not so surprising in the light of Schoor's algorithm.

\section{Conclusions}

We have presented a biclique-based approach for compressing and multiplying Boolean matrices, aimed at reducing both space usage and computational cost in large-scale graph operations. By exploiting dense bipartite substructures, our representation encodes sets of edges implicitly while maintaining compatibility with algebraic operations.

In order to obtain large bicliques efficiently, we improved our previous biclique extraction process with a percentile-guided adaptive mechanism that dynamically tunes its parameters based on the degree distribution of the input graph. This optimization significantly improved compression efficiency and reduced extraction time compared to our previous implementation.

On our representation, we extended the Schoor algorithm \cite{Schoor82} used by Arroyuelo et al.~\cite{arroyuelo2025evaluating}, so as to operate directly on biclique-encoded matrices, decomposing the product into four subcomponents---matrix--matrix, matrix--biclique, biclique--matrix, and biclique--biclique multiplications. The compositional nature of this formulation allows the result to be represented as a new set of bicliques, thereby operating always on the compressed domain.

Experimental results on web and social graphs demonstrated that our method achieves up to 290 times times faster multiplication while using up to 10 times less space than Arroyuelo et al.'s \textit{CSR}/\textit{CSC} representation. The biclique representation shows to provide even better compression on the matrices resulting from multiplications, and provides a promising foundation for scalable algebraic computations on massive sparse graphs.

Future work includes exploring parallel and GPU implementations of the biclique-based multiplication, integrating the method with graph query engines, and extending the model to weighted and dynamic graphs where biclique structures evolve over time. We also plan to explore whether transitive closure algorithms implemented on \textit{CSR}/\textit{CSC} \cite{arroyuelo2025evaluating} can be accelerated with the biclique-based representation, in order to complete a full Boolean algebra.

\section*{Acknowledgements}

CH was funded by ANID Fondecyt Grant 11240971; GN was funded by ANID -- Millennium Science Initiative Program -- Code ICN17\_002, and Fondecyt Grant 1-230755, ANID, Chile.

\bibliographystyle{elsarticle-num}
\bibliography{references}

\end{document}